\documentclass[11pt,a4paper]{article}
\usepackage{bm}
\usepackage{amssymb}
\usepackage{amsmath,amsthm}
\usepackage{epsfig}
\usepackage{slashed}
\usepackage{authblk}
\usepackage{hhline}
\usepackage{color}
\usepackage{multirow}
\usepackage[normalem]{ulem}
\usepackage[T1]{fontenc}   

\usepackage{graphics}
\usepackage{easybmat}
\usepackage{epsfig,float,amsmath,amsfonts,amssymb,bm,bbm,array,dcolumn,verbatim,mathrsfs,subfigure}

\usepackage{amsmath, amssymb, amsfonts,lipsum}
\usepackage{hyperref}
\usepackage{graphicx, subfigure, epstopdf}
\usepackage{color}

\def\be{\begin{equation}}
\def\ee{\end{equation}}

\def\bea{\begin{eqnarray}}
\def\eea{\end{eqnarray}}
\begin{document}

\title{On Electroweak Phase Transition and Di-photon Excess with a 750 GeV Scalar Resonance}
\author{Anish Ghoshal}
\affil{\small{\textit{Department of Physics, Indian Institute of Technology-Madras, Chennai-600036}}}
\maketitle

\begin{abstract}
For successful electroweak baryogenesis to take place through the sphaleron process the universe needs to undergo a strong
first order cosmological phase transition. While it does not occur in the Standard Model it becomes possible in the presence of extra scalars
in BSM. One of these scalars can well be responsible for the recently observed diphoton excess in the CMS and ATLAS experiments in LHC.
We study the Electroweak phase transition in a myriad of scalar models in this context.
\end{abstract}

\section{Introduction}
The cosmological matter-antimatter asymmetry ( $ Y_B = (8.58 ± 0.22)\times 10^{−11} ) $ [BAU]\cite{Ade:2013sjv} may have been generated at temperatures much higher than the electroweak scale, e.g.
heavy Majorana neutrino decay \cite{Pilaftsis:1997jf,Flanz:1996fb,Pilaftsis:1998pd} or even by squarks \cite{Huet:1995sh}. These models are difficult to test. Electroweak baryogenesis, on the other hand, may be governed by
the physics that is currently being probed at the LHC experiment and will continue to be so at future particle colliders.

The initial conditions assumed for EWBG are a hot, radiation-dominated early universe containing zero net baryon charge in which full $ SU(2)_{L} \times U(1)_{Y}$ electroweak symmetry is manifest. As the temperature cools below the critical temperature the higgs field settles down to a vacuum state and spontaneously breaks the EW symmetry. During this transition EWBG occurs.
The standard Model of particle physics itself contains the necessary ingredients \cite{Kuzmin:1985mm,Rubakov:1996vz,Klinkhamer:1984di,Arnold:1987mh} for successful electroweak baryogenesis but the parameters
of the model is not compatible with the Sakharov conditions \cite{Sakharov:1967dj}. Therefore Beyond Standard Model physics is required. With this empirically based motivation to look for physics beyond the Standard Model it may be natural to
extend the scalar sector in context with the electroweak phase transition. And this scenario receives a further boost as after finding the standard model
scalar Higgs at the LHC \cite{Aad:2012tfa,Chatrchyan:2012xdj} recently di-photon excess was observed at CMS and ATLAS \cite{atlas,CMS:2015dxe} in LHC with an invariant mass of 750 GeV.
Previously studied models of EW baryogenesis include BSM with scalar singlets \cite{Choi:1993cv,Espinosa:1993bs,Ham:2004cf,Cline:2009sn,Ahriche:2007jp,Profumo:2007wc}, scalar doublets\cite{Shu:2013uua,Cline:2011mm,Huber:2006ik,Chao:2014dpa,Dorsch:2013wja,Cline:2013bln,Gil:2012ya,Blinov:2015vma,Racker:2013lua,Borah:2012pu}, scalar triplets \cite{Cohen:2012zza,Huang:2015izx}. It was shown that the strength of the EW phase transition could be strengthened by these extensions. Similar study is also done in MSSM and
its extensions\cite{Cline:1998hy}. The diphoton excess has been explained in a plethora of supersymmetric \cite{Ding:2015rxx}
and non-supersymmetric models \cite{Becirevic:2015fmu,Han:2015qqj,Backovic:2015fnp}. In this article we try to study some of the models which can suitably explain EW
phase transition and the di-photon excess simultaneously. The paper is organized as follows: Section II considers a extending the
SM by a real singlet with a 750 GeV mass. Section III introduces a 2HDM model with four scalars, Section IV presents a BSM scalar with color quantum
number. In Section V we discuss the process of electroweak phase transition after accounting for the finite temperature and loop-level correction of the
scalar potential. Next section describes the observation of di-photon excess in CMS and ATLAS. Section VII presents some elementary results of the study on these models. Next we discuss some of the other possible models. We make some model-independent remarks on the scalar models in context to phase transition in section IX and conclude with general comments in section X.

\section{Scalar Singlet}
Tree-level potential for the Higgs doublet H and a real singlet scalar S is given by similar to \cite{Chowdhury:2011ga}
\begin{eqnarray}
	V_0 &=&  \lambda_h\left(|H|^2 -\frac12 v_0^2\right)^2 +
	 \frac14\lambda_s\left( S^2 - w_0^2\right)^2 
	 \nonumber\\
	 &+& \frac12\lambda_m |H|^2 S^2 \,.
\end{eqnarray}
This potential has the $Z_2$ symmetry but is broken spontaneously at high temperatures, giving $S$ a VEV (with $H=0$) in the electroweak symmetric vacuum.The true vacuum is along the $H$ axis at $T=0$. The real fields become $H=h/\sqrt{2}$ and $S$.

\section{2HDM}
The most general gauge-invariant and renormalizable potential that can be written for two Higgs doublets, allowing for a soft breaking of the Z
\label{2HDM_potential}
	\begin{equation}	
	\label{2HDM_potential}
	\begin{split}
		V_{\rm tree}(H_1,H_2)=&-\mu^2_1 H_1^{\dagger}H_1-
					  \mu^2_2H_2^{\dagger}H_2-
					  \frac{\mu^2}{2}\left(e^{i\phi}H_1^{\dagger}H_2+h.c.\right)+\\
				 &+\frac{\lambda_1}{2}\left(H_1^{\dagger}H_1\right)^2+
				   \frac{\lambda_2}{2}\left(H_2^{\dagger}H_2\right)^2+
				   \lambda_3\left(H_1^{\dagger}H_1\right)\left(H_2^{\dagger}H_2\right)+\\
				 &+\lambda_4\left(H_1^{\dagger}H_2\right)\left(H_2^{\dagger}H_1\right)+
			 	   \frac{\lambda_5}{2}\left[\left(H_1^{\dagger}H_2\right)^2+h.c.\right],
	\end{split}
\end{equation}
where we make $\lambda_5$ real by redefining the fields. The 2HDM scalar potential,
Eq. (2), can violate CP either explicitly, via the complex phase $\phi$ in the term $(H_1^{\dagger}H_2+h.c.)$. For matter-antimatter asymmetry, the BSM
sources of extra CP violation are important in the context of generating the
observed BAU. Here we are interested in the nature of the electroweak phase
transition only, so we will restrict ourselves to a CP conserving scalar sector
to make analysis. Also from \cite{Fromme:2006cm,Cline:2011mm}, it can be understood that the phase does
not affect the phase transition significantly.
The doublets and their VEVs at electroweak minimum can be written as
\begin{equation*}
	H_{i} = \left( \begin{array}{c} \varphi_i^+ \\ h_i+i\eta_i \end{array} \right),
\end{equation*}

\begin{equation}
	 \label{EWmin}
	\langle H_1\rangle = \left( \begin{array}{c} 0 \\ v\cos\beta \end{array} \right),\hspace{0.5cm}
  	\langle H_2\rangle = \left( \begin{array}{c} 0 \\ v\sin\beta \end{array} \right),
\end{equation}
with $v=246/\sqrt{2}$~GeV. The parameter $\beta$, gives the mixing in yhe form of a change of basis:
\[\begin{array}{l} H_1^\prime=\cos\beta\ H_1+\sin\beta\ H_2\\
   H_2^\prime=-\sin\beta\ H_1+\cos\beta\ H_2 \end{array} \implies
   \langle H_1^\prime\rangle= \left( \begin{array}{c} 0\\ v \end{array} \right) \text{ \ and \ } 
\langle H_2^\prime\rangle=0 \]
it becomes clear that $H_1^\prime$ behaves like the SM doublet, therefore its upper component must be the charged 
Goldstone boson ($G^+$) and the lower component contains the neutral Goldstone ($G^0$). Similar to $\beta$ as the mixing angle we define $\alpha$ to be the 
mixing angle between the lightest and heaviest CP-even fields, denoted $h^0$ and $H^0$. The physical states are
\begin{alignat*}{2}
	&G^+=\cos\beta\ \varphi_1^+ + \sin\beta\ \varphi_2^+ && \hspace{1cm} \text{(charged Goldstone),}\\
	&H^+=-\sin\beta\ \varphi_1^+ + \cos\beta\ \varphi_2^+ && \hspace{1cm} \text{(charged Higgs),}\\
	&G^0=\cos\beta\ \eta_1 + \sin\beta\ \eta_2 && \hspace{1cm} \text{(neutral Goldstone),}\\
	&A^0=-\sin\beta\ \eta_1 + \cos\beta\ \eta_2 && \hspace{1cm} \text{(CP-odd Higgs),}\\
	&h^0=\cos\alpha\ h_1 + \sin\alpha\ h_2 && \hspace{1cm} \text{(lightest CP-even Higgs)},\\
	&H^0=-\sin\alpha\ h_1 + \cos\alpha\ h_2 && \hspace{1cm} \text{(heaviest CP-even Higgs)}.
\end{alignat*}
The SM-like Higgs is an admixture of $h^{0}$ and $H^{0}$. But had their been a
phase, that would have entered in the mixing causing CP-violating effects.
That been set to zero simplifies the problem a lot.
The condition that Eq. (3) be a minimum can gives us:
\begin{equation}
	\label{As}\begin{split}
	&\mu_1^2= v^2\left(\lambda_1\cos^2\beta+\lambda_{345}\sin^2\beta\right)-
			    M^2\sin^2\beta,\\
	&\mu_2^2=v^2\left(\lambda_2\sin^2\beta+\lambda_{345}\cos^2\beta\right)-
			   M^2\cos^2\beta,
\end{split}\end{equation}
where $M^2\equiv \mu^2/\sin(2\beta)$ and $\lambda_{345}\equiv\lambda_3+\lambda_4+\lambda_5$. The parameter $M$ plays the role of a natural scale for the masses of the additional scalars, while $h_{SM}$ scales with $v$ as usual. From the diagonalization of the mass matrix we also see that the quartic couplings can be written in terms of the physical parameters as
\begin{equation}\begin{split}
	\label{couplings}
	&\lambda_1=\frac{1}{2v^2\cos^2\beta} 		
		  \left(m_{h^0}^2\cos^2\alpha+m_{H^0}^2\sin^2\alpha-M^2\sin^2\beta\right),\\
	&\lambda_2=\frac{1}{2v^2\sin^2\beta}
		   \left(m_{h^0}^2\sin^2\alpha+m_{H^0}^2\cos^2\alpha-M^2\cos^2\beta\right),\\
	&\lambda_3=\frac{1}{2v^2\sin(2\beta)}
		   \Big[\left(2m_{H^\pm}^2-M^2\right)\sin(2\beta)-\left(m_{H^0}^2-m_{h^0}^2\right)\sin(2\alpha)\Big],\\
	&\lambda_4+\lambda_5=\frac{1}{v^2}\left(M^2-m_{H^\pm}^2\right),\\
	&\lambda_4-\lambda_5=\frac{1}{v^2}\left(m_{A^0}^2-m_{H^\pm}^2\right).
\end{split}\end{equation}
These relations allow us to express the coefficients in the potential as a function of the masses ($m_{h^0}$, $m_{H^0}$, $m_{A^0}$, $m_{H^\pm}$) and the mixing angles ($\beta$, $\alpha$) which now become the input parameters of the model.

\section{Scalar with Color}
A new colored scalar $X_{c}$ similar to \cite{Cohen:2011ap} with hypercharge 1/3 and a bare mass term $M_X$ is added to the SM. It
couples to the Higgs boson via a quartic coupling, $ \Lambda $ :
\begin{equation}
\mathcal{L} \supset -M_X^2 |X_c|^2 - \Lambda |X_c|^2 |H|^2 - \frac{K}{6} |X_c|^4,
\end{equation}
where $K$ is the quartic self coupling for $X_c$, and $H$ is the Higgs doublet.

\section{Electroweak Phase Transition}
The effective potential at finite temperature T can be written as
\begin{equation}
	V = V_{\rm tree} + V_{\rm loop}+V_{CT} + V_T.
\end{equation}
where $V_{\rm tree}$ , $V_{\rm loop}$ and $V_T$ are tree-level, one-loop temperature-independent
and -dependent pieces, respectively. The tree-level potential $V_{\rm tree}$ has been
given in the models discussed above. Working in the Landau gauge ($\xi = 0$),
the temperature-independent one-loop correction has the form \cite{Dolan:1973qd}
\begin{equation}
 V_1 = \sum_i \frac{n_i}{64\pi^2} m^4_i(v,\phi) \left(\ln \frac{m^2_i(v,\phi)} {Q^2} - C_i \right).
\end{equation}
with $i$ indexing the particles summed over all the particles coupling to the doublets and $n_i$ their numbers of degrees of freedom. 
(positive for bosons and negative for fermions as they contribute positively and negative to the bosonic and fermionic loops
respectively. $C_i$ are renormalization-scheme-dependent constants 
($C_i = 1/2$ for transverse gauge bosons and $3/2$ for 
the rest in the $\overline{\mathrm{MS}}$ scheme); $m^2_i(v,\phi)$ is the field-dependent squared mass for each species.
The one loop corrections are dominated by particles of high mass, namely the W,Z bosons and the top quarks. As well as the new scalar(s).
The above expression is not renormalized. We impose the condition $\textit{v}$ = 246.22 Gev and $m_h = 125 $ Gev. The counter-terms are absorbed in $V_{loop}$. With $\frac{dV_1}{d\phi}\big|_{\phi=v}=0$ and $\frac{d^2V_1}{d\phi^2}\big|_{\phi=v}=0$, we get one loop correction
\begin{eqnarray}\label{renormalizedV1}
\bar{V}_{1,i}&=&\pm\frac{n_i}{64\pi^2}\Bigg(\big(m^2_i(\phi)\big)^2\ln m^2_i(\phi)+\bigg[\Big(-\frac{3}{4}\frac{m^2_i(v)m^2_i(v)'}{v}+\frac{3}{4}\big(m^2_i(v)'\big)^2+\frac{1}{4}m^2_i(v)m^2_i(v)''\Big)\nonumber\\
&&+\Big(-\frac{3}{2}\frac{m^2_i(v)m^2_i(v)'}{v}+\frac{1}{2}\big(m^2_i(v)'\big)^2+\frac{1}{2}m^2_i(v)m^2_i(v)''\Big)\ln m^2_i(v)\bigg]\phi^2\nonumber\\
&&+\bigg[\Big(\frac{1}{8}\frac{m^2_i(v)m^2_i(v)'}{v}-\frac{3}{8}\big(m^2_i(v)'\big)^2-\frac{1}{8}m^2_i(v)m^2_i(v)''\Big)\nonumber\\
&&+\Big(\frac{1}{4}\frac{m^2_i(v)m^2_i(v)'}{v}-\frac{1}{4}\big(m^2_i(v)'\big)^2-\frac{1}{4}m^2_i(v)m^2_i(v)''\Big)\ln m^2_i(v)\bigg]\frac{\phi^4}{v^2}\Bigg)\\
&\to&\pm\frac{n_i}{64\pi^2}\bigg(\big(m_i^2(\phi)\big)^2\left(\ln\frac{m_i^2(\phi)}{m_i^2(v)}-\frac{3}{2}\right)+2m_i^2(v)m_i^2(\phi)\bigg).
\end{eqnarray}
The first equation is for a generic mass square as a function of $\phi^2$ so that the counter term in the effective potential is up to $\phi^4$, while the second one has $m^2_i(\phi)=a+b\phi^2$. Only if there's a BSM fermion postulated it will follow the first equation. Rest of the particle species will follow the first equation.
The 1-loop thermal corrections to the effective potential are given by~\cite{RubakovGourbunov}
\begin{equation}
	\label{Vthermal}
	V_{\rm thermal}=\frac{T^4}{2\pi^2}\sum_i n_i\int_0^{\infty} x^2\text{ln}\left(1\mp 
e^{-\sqrt{x^2+m_i^2/T^2}}\right)dx,
\end{equation}
where the sign inside the logarithm is $-$ for bosons and $+$ for fermions as bosonic and fermionic loops contribute positively an negatively respectively. We use an approximation for the integral and use MATHEMATICA notebook to compute further. At high temperatures, the equation can be approximated by 
\begin{equation}
\begin{split}
	V^{HT}_{\rm thermal}&\approx T^4\sum_B n_B\left[-\frac{\pi^2}{90}+
						\frac{1}{24}\left(\frac{m_B}{T}\right)^2
						-\frac{1}{12\pi}\left(\frac{m_B}{T}\right)^3
						-\frac{1}{64\pi^2}\left(\frac{m_B}{T}\right)^4\text{ln}\frac{m_B^2}{c_BT^2}\right]\\
			   &+T^4\sum_F n_F\left[-\frac{7\pi^2}{720}
					+\frac{1}{48}\left(\frac{m_F}{T}\right)^2
					+\frac{1}{64\pi^2}\left(\frac{m_F}{T}\right)^4\text{ln}\frac{m_F^2}{c_FT^2}\right]\\
\end{split}\end{equation}
with $c_F=\pi^2\text{exp}\left(\frac{3}{2}-2\gamma\right)$ and $c_B=16c_F$, and at low temperatures
\begin{equation}
	V_{\rm thermal}^{LT}\approx-T^4\sum_{i=B,F}n_i\left(\frac{m_i}{2\pi T}\right)^{3/2} \exp\left(-\frac{m_i}{T}\right)\left(1+\frac{15}{8}\frac{T}{m_i}\right).
\end{equation} 
Although there will be deviation from the actual result due to this approximation but that is small compared to rest. Moreover this study is only an elementary-level one and we would continue to use more rigorous mechanisms to study the model in future.

And the scalar masses from resummation of daisy diagrams: 
diagrams~\cite{Carrington:1991hz}. The mass-matrix
\begin{equation}
	(M_T)_{ij}=\frac{1}{2}\frac{\partial^2 }{\partial\phi_i\partial\phi_j}
		\left(V_{tree}+\frac{T^2}{24}\sum_i n_i m_i^2\right).
\end{equation}
The thermally corrected masses are then the eigenvalues of $M_T$. Ultimately the potential becomes:
\begin{equation}
	V=V_{\rm tree}+V_{\rm loop}+V_{CT}+V_T.
\end{equation}

Usually $m^2\propto\phi^2$, so the $\frac{n_iT^4}{2\pi^2}\times\frac{\pi^2}{12}\frac{m^2}{T^2}$ ($\frac{\pi^2}{24}\frac{m^2}{T^2}$) term is also quadratic in $\phi$, and adds to the tree level negative Higgs mass term $-\frac{1}{4}m_h^2\phi^2$ a positive contribution proportional to $T^2\phi^2$ at high temperature, making the quadratic term positive. Let the potential be
\begin{equation}\label{thermalMassAndCubic}
V=d(T^2-T_0^2)\phi^2-eT\phi^3+\frac{1}{4}\lambda\phi^4,
\end{equation}
with correction from thermal effect included. The contribution of the cubic term to the Electroweak phase transition is palpable:
\begin{equation}\label{SMStrength}
\frac{\langle\phi(T_c)\rangle}{T_c}=\frac{2e}{\lambda}.
\end{equation}
Here $T_c$ is the critical temperature where the potential at the new local minimum $\phi\neq0$ is equal to the potential at $\phi=0$. The nucleation and bubble formation is usually right after the universe goes below the critical temperature, so we take $T_c\simeq T_n$ . In the SM $e=\frac{1}{6\pi v^3}(2m_W^3+m_Z^3)=6.4\times10^{-3}$ \cite{Dine:1992vs}, so the phase transition strength for $m_h\simeq125$~GeV is $\frac{2e}{\lambda}=0.1$, is weak for a successful EW baryogenesis.
Eq.~(12) suggests a straightforward way to strengthen the phase transition: add new strongly interacting bosons, which is exactly the new boson(s) in the model(s). If we simply ignore the term $m_s$ and put $m_i^2=\frac{1}{2}y^2\phi^2$, the cubic term in $\phi$ in the expansion for bosons as part of $e$ term, for a strongly first order phase transition the new boson only need to contribute $n_iy^3\simeq6$, which implies $y\simeq1$. This nonetheless is not a good approximation. A better way lies in the above bosonic high temperature expansion Eq.~(12), by comparing the quadratic term in the expansion to the tree level mass term. The SM particle thermal masses are well known. As for the new bosons, we already have $m_s$ of order hundred of~GeVs even at Higgs VEV $\phi=0$ just like thermal mass, and the thermal mass itself of the order $y^2T^2$, we expect its effect to be subdued.

Below the critical temperature $T_{c}$, the minimum breaks electroweak symmetry. But as
the global minimum of the potential, the field is still in the symmetric local minimum
because the two minima are separated by a potential barrier due to thermal
fluctuations. The transition proceeds via thermal
tunneling, as the quantum tunneling probability has decreased due to the thermal barrier. In
the broken phase (with non-zero vev) of the symmetric background, after nucleation the
bubbles grow, it converts a false vacuum into true one. The whole Universe transitions
into this broken phase.
The sphalerons are suppressed in the broken phase due to the W bosons gaining
mass, and the weak interactions act only on very short distance scales. Hence, this suppression
is proportional to the gauge boson masses and proportional to the Higgs vev right after
the phase transition. 
A simple criterion for sphaleron freeze-out is obtained by assuming
that the sphaleron processes decouple when their rate becomes smaller than the expansion of the U and this condition is given by $\frac{\textit{$v_{c}$}}{T_{c}}  \geq 1 $

A very good account of EWPT can be found in \cite{RubakovGourbunov} and \cite{Chung:2012vg}.

\section{Di-photon Excess}
Both ATLAS and CMS collaborations recently announced a search for resonances in the di-photon channel, featuring an excess of events around $m_{\gamma \gamma}\approx$ 750 GeV \cite{atlas, CMS:2015dxe}.

\begin{align}
&\sigma(pp\to S)\times  \text{BR}(S\to\gamma\gamma) \gtrsim 2 \text{ fb}\label{feature1}\ , \\
&m_S \approx 750 \text{ GeV}\label{feature2},
\end{align}
where $S$ is the new resonance singly produced in $pp$ collisions.
The uncanny feature of this di-photon excess is a large total width of the resonance. In terms of events, ATLAS collaboration has a better fit and favors a rather large width of resonance around 6 per cent of the resonance mass. Here we assume 
\begin{equation}
\frac{\Gamma_{\mathrm{tot}}(S)}{m_S}\approx \text{3-9} \% \ . \label{feature3}
\end{equation}
However there are no excesses for the dijet \cite{Khachatryan:2015dcf}, $t\bar{t}$ \cite{Khachatryan:2015sma}, diboson or dilepton channels. Which leaves us at a challenging front to account for the 750 GeV resonance. 
The 2HDM model alone cannot produce the large cross-section observed. So it has been extended by adding a new vector-like fermion \cite{Angelescu:2015uiz} and other scalars \cite{Han:2015qqj}

One may attempt to interpret the di-photon excess at as the resonant production of the singlet scalar $S$ with mass $M_S = 750$~GeV. Considering the possible production mechanisms for the resonance at 750~GeV it is important to note that the CMS and ATLAS did not report a signal in the $\sim 20  {\rm{fb}}^{-1}$ data at 8~TeV in Run 1. This may be because 750~GeV resonance is produced through a mechanism with a steeper energy dependence, e.g. gluon-gluon fusion. And as the centre-of-mass energy increases in the run the parton contribution of gluons and heavier quarks increases. Here for a simple analysis we consider this as the dominant production mechanism and no other associated production. So the scalar with mass 750~GeV decays to two photons via a loop. The cross section can be expressed as
\begin{align}
	\label{eq:cs}
	\sigma(pp \to \gamma \gamma) = \frac{C_{gg}}{M_S s} 
	\Gamma_{gg} \text{Br}_{\gamma\gamma},
\end{align}
with the proton centre-of-mass energy $\sqrt{s}$ and the parton distribution integral $C_{gg} = 174$ at $\sqrt{s} = 8$~TeV and $C_{gg} = 2137$ at $\sqrt{s} = 13$~TeV \cite{Franceschini:2015kwy}. One can obtain a best fit guess of the cross section by reconstructing the likelihood, assumed to be Gaussian, from the $95\%$ C.L. expected and observed limits in an experimental search. For the diphoton excess, we use a best fit cross section value of 7~fb found by combining the 95\% CL ranges from ATLAS and CMS at 13~TeV and 8~TeV for a resonance mass of 750~GeV. 

Apart from the necessary decay modes of the scalar $S$ i.e, $S\to gg$ and $S\to \gamma\gamma$, $S$ may also decay to other particles; due to SM invariance and the fact that $M_S > m_Z$, $S\to \gamma\gamma$ necessitates the decays $S\to \gamma Z$ and $Z Z$ which are suppressed by $2\tan^2 \theta_W \approx 0.6$ and $\tan^4\theta_W \approx 0.1$ relative to $\Gamma(S \to \gamma\gamma)$. $S$ may also decay to SM fermions due to mixing with the heavy vector-like fermions which in turn affects the phase transition. The total width is thereby given by $\Gamma_S \approx \Gamma_{gg} + 1.7\times\Gamma_{\gamma\gamma} + \Gamma_{t\bar{t}}$. \cite{Deppisch:2016scs}

Production of a scalar resonance in gluon fusion via a loop of top and a vector-like quark and subsequent decay of scalar resonance to $\gamma\gamma$ via a loop of vector-like quark and the charged scalar in 2HDM is considered. These fermions should have vector-like couplings to the electroweak gauge bosons in order to avoid generating
their masses through the Higgs mechanism only and then be compatible with the electroweak
precision tests as well as the LHC Higgs data. Vector-like fermions appear in many
extensions of the SM and recent discussions have been given in Ref. \cite{Ellis:2014dza}.
To fulfill SM gauge invariance, at least two vector-like multiplets need to be introduced in order to generate directly
Yukawa couplings for the VLQ that are not suppressed by SMQ-VLQ mixing angles. Another fine-tuning arrangement can be made in the lines of vector-like leptons \cite{Angelescu:2015uiz}. But here, for simplicity we consider a singlet.
The partial decay widths are given by \cite{McDermott:2015sck}
\begin{align}
\label{3.8}
	\Gamma_{\gamma\gamma} &= \phantom{K}
		\frac{\alpha^2\, M^3_S}{256 \pi^3} 
    \left| \sum_F
			\frac{N^C_F Q^2_F \lambda'_{SFF}}{M_F} 
			\mathcal{A}\left(\frac{m_S^2}{4 M_F^2} \right) 
		\right|^2, \nonumber\\
	\Gamma_{gg} &=
		K \frac{\alpha^2_s M^3_S}{128 \pi^3} 
		\left| \sum_F^C
			\frac{\lambda'_{SFF}}{M_F} 
			\mathcal{A}\left(\frac{m_S^2}{4 M_F^2} \right)
		\right|^2.
\end{align}
Here, the sums in $\Gamma_{\gamma\gamma}$ and $\Gamma_{gg}$ are over colored fermion species and flavors in the first case and all species with electric charges in the second. $N_F^C$ is the number of color degrees of freedom of a species, i.e  3 for vector-like quark. Similarly, $Q_F$ is the electric charge of the species. The coupling of $S$ to a fermion species is $\lambda'_{SFF}$. $A(x)$ is a loop function defined by
\begin{align}
\label{3.9}
	A(x) = \frac{2}{x^2} [x + (x - 1) f(x)],
\end{align}
with
\begin{align}
\label{3.10}
	f(x) = \left\{ 
		\begin{matrix}
			\arcsin^{2} \sqrt{x} & & x \leq 1 \\ 
	    -\frac{1}{4}\left[\ln\left(\frac{1+\sqrt{1-x}}{1-\sqrt{1-x}}\right)-i\pi\right]^2 
			& & x > 1.
		\end{matrix} 
	\right.
\end{align}

In addition, the decay of $S$ to a pair of fermions is given by
\begin{align}
\label{3.11}
	\Gamma_{f\bar{f}} = \frac{N_f^C \lambda'^2_{Fff} M_S}{16\pi}
	\left(1 - \frac{4 M_F^2}{M_S^2} \right)^{2/3}
\end{align}

As discussed in \cite{Franceschini:2015kwy} in a model-independent manner, the di-photon excess can be explained for $10^{-6} \lesssim \Gamma_{gg}/M_S \lesssim 2 \times 10^{-3}$ (the upper limit from dijet searches) and $\Gamma_{\gamma\gamma}/M_S \approx 10^{-6}$, as long as $gg$ and $\gamma\gamma$ are the only decay modes of $S$.
For a full analysis one needs to use:
The signal cross section at proton centre-of-mass energy $\sqrt{s}$ (= 8 or 13 TeV)
\begin{equation}
 \sigma(pp \to S \to \gamma\gamma) =\frac{2J+1}{M\Gamma s} \bigg[C_{gg} \Gamma(S\to gg) +\sum_q C_{q\bar q} \Gamma(S\to q\bar q)\bigg]\Gamma(S\to \gamma\gamma) \, ,
\label{eq:sigmasig}
\end{equation}
where the relevant decay $S$ widths are evaluated at leading order in QCD. The $2J+1$ factor could be reabsorbed by redefining the widths as summed over all $S$ components. The decay into two photons implies that the two relevant cases are $J=0,2$.
We take spin-0 resonance and pdf as in \cite{Franceschini:2015kwy}.

For the CP-even Higgs decay into $\gamma\gamma$ \cite{Djouadi:2005gj}
\be \label{hrrgs}
    \Gamma(S \to \gamma\gamma)  =   \displaystyle
            \frac{\alpha^2 m_S^3}{256 \pi^3 v^2}
            \left| \sum_i y_i N_{ci} Q_i^2 F_i \right|^2,
    \ee
with $N_{ci}$, $Q_i$ are the color factor and the electric charge
respectively for particles running in the loop. The dimensionless
loop factors for particles of spin given in the subscript are \be
    F_1(\tau) = 2 + 3 \tau + 3\tau (2-\tau) f(\tau),\quad
    F_{1/2}(\tau) = -2\tau [1 + (1-\tau)f(\tau)],\quad
    F_0(\tau) = \tau [1 - \tau f(\tau)],
\ee where $\tau=4m_i^2/m_S^2$ and $y_i$ are the Yukawa couplings of the particles running in the loops as shown in fig. \ and 
\begin{equation}
    f(\tau) = \left\{ \begin{array}{lr}
        [\sin^{-1}(1/\sqrt{\tau})]^2, & \tau \geq 1 \\
        -\frac{1}{4} [\ln(\eta_+/\eta_-) - i \pi]^2, & \, \tau < 1
        \end{array}  \right.\label{hggf12}
\end{equation}

\begin{figure}[H]
\centering
\includegraphics[width=80mm]{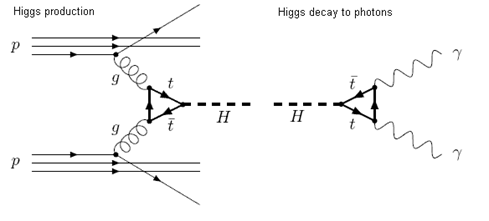}
	\caption{\footnotesize Feynman diagram showing the production of di-photons from gluon-gluon fusion formation of the SM Higgs. In the same diagram if we substitute Higgs with the 750 GeV scalar and 't' with the new vector-like fermion we get the diagram we study.}\label{}
\end{figure}

\section{Results}
\subsection{Scalar Singlet}

Expressing the masses of the scalar in terms of the couplings and using the $\lambda_{m}$ and critical vev field values as input parameters a scan was done($\textit{$v_{c}$}$ 0 to 300 GeV and $\textit{$w_{c}$}$ over the range 0 to 300 GeV). When $\lambda_{m}$ was kept less than unity no parameter space suiting $\xi_{c}$ greater than equal to 1 was found. However on increasing the bound on $\lambda_{m}$ to 4$\pi$ the following plot was obtained. Mass of the second scalar was put as constrained to be 250-1500 GeV. 
\begin{figure}[H]
\centering
\includegraphics[width=80mm]{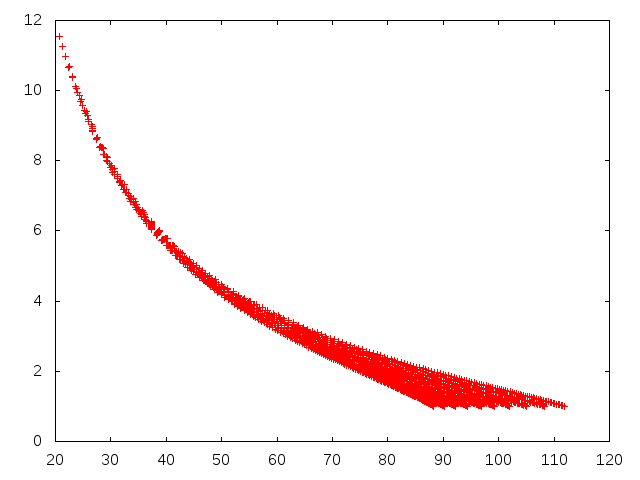}
	\caption{Plot of $\xi_{c}$(Y-axis) vs $T_c$(X-axis)}\label{}
\end{figure}

Next the mass of the second scalar $m_{s}$ was restricted to be 740-760 GeV. Then also no parameter space was obtained. This goes on to show the new inert scalar cannot satisfy the criterion of the 750 GeV resonance as well as be responsible for EW phase transition and dark matter as was described in \cite{Cline:2012hg}.
[We also put in the constraint that the EW minima should be less than the other one.]
\par The electroweak physics of such a scalar is only governed by its mixing angle with the Higgs. And so the opening up of parameter relies heavily on $\lambda_{m}$. Such a scalar along with a vector-like fermion can accommodate the di-photon excess \cite{Falkowski:2015swt}. This suggests the critical temperature to be low(which is not so desirable) for stronger phase transition. However any $\xi_{c}$ grater than equal to 1 can satisfy the condition. So a region in the plot where the critical temperature is not too small may explain EWPT.

\subsection{2HDM}
2HDM have been studied in context with collider phenomenology and the parameters are constrained from their results. We add a vector-like fermion having electric charge of (2/3)units and color. This is done keeping in mind to explain the observed enhanced cross-section in the di-photon excess.
From the plots of \cite{Becirevic:2015fmu} 
\begin{figure}[H]
\centering
\includegraphics[width=80mm]{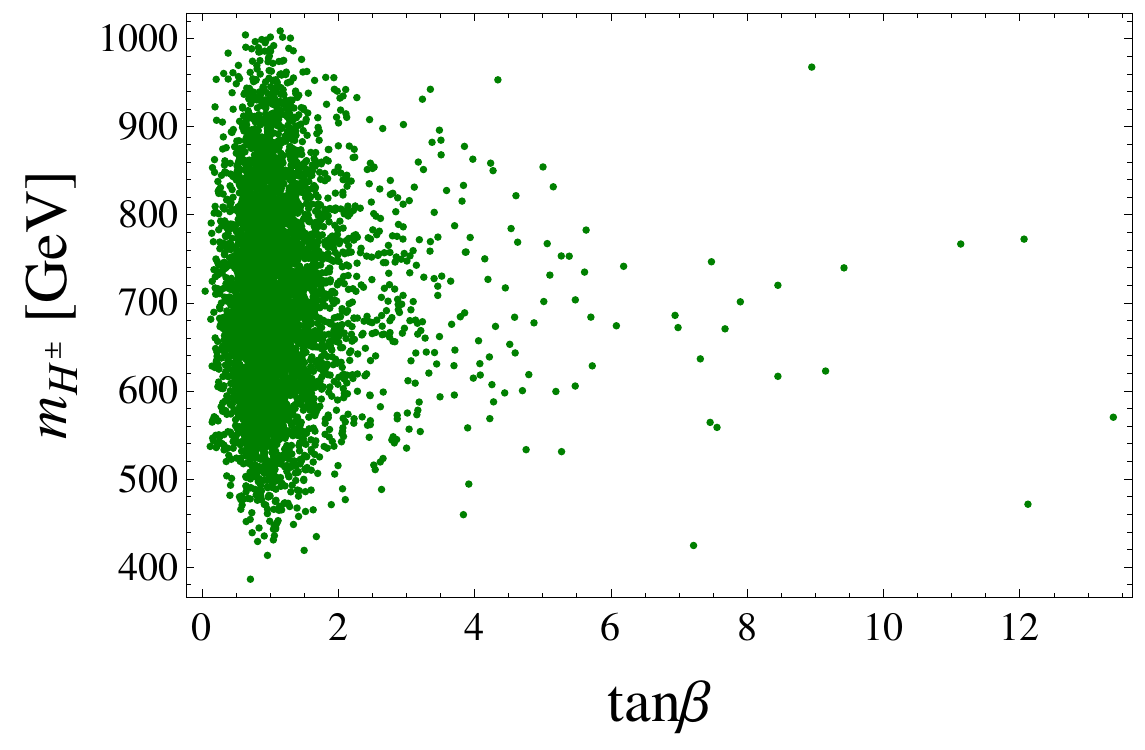}
\caption{ {\footnotesize Points in parameter space allowed. Mass of the charged scalar in Y-axis and the mixing angle tan$\beta$ in X-axis. } \label{fig1} }
\end{figure}
We take the constraint tan$\beta$ < 5. Rest of the input parameter masses are set to 1 TeV.
We get the following plots:

\begin{figure}[H]
\centering
 \includegraphics[width=70mm,height=50mm]{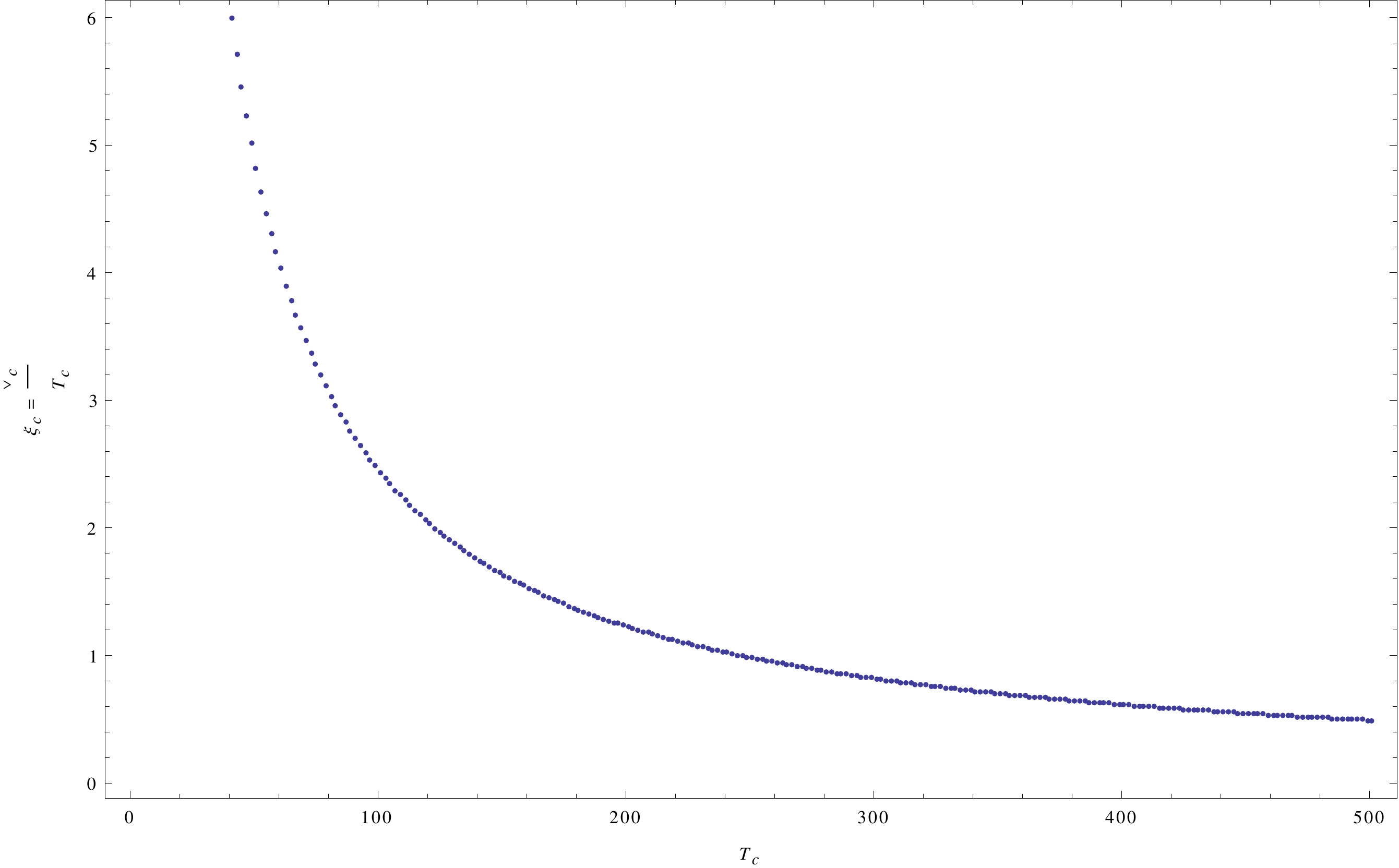}
 \includegraphics[width=70mm,height=50mm]{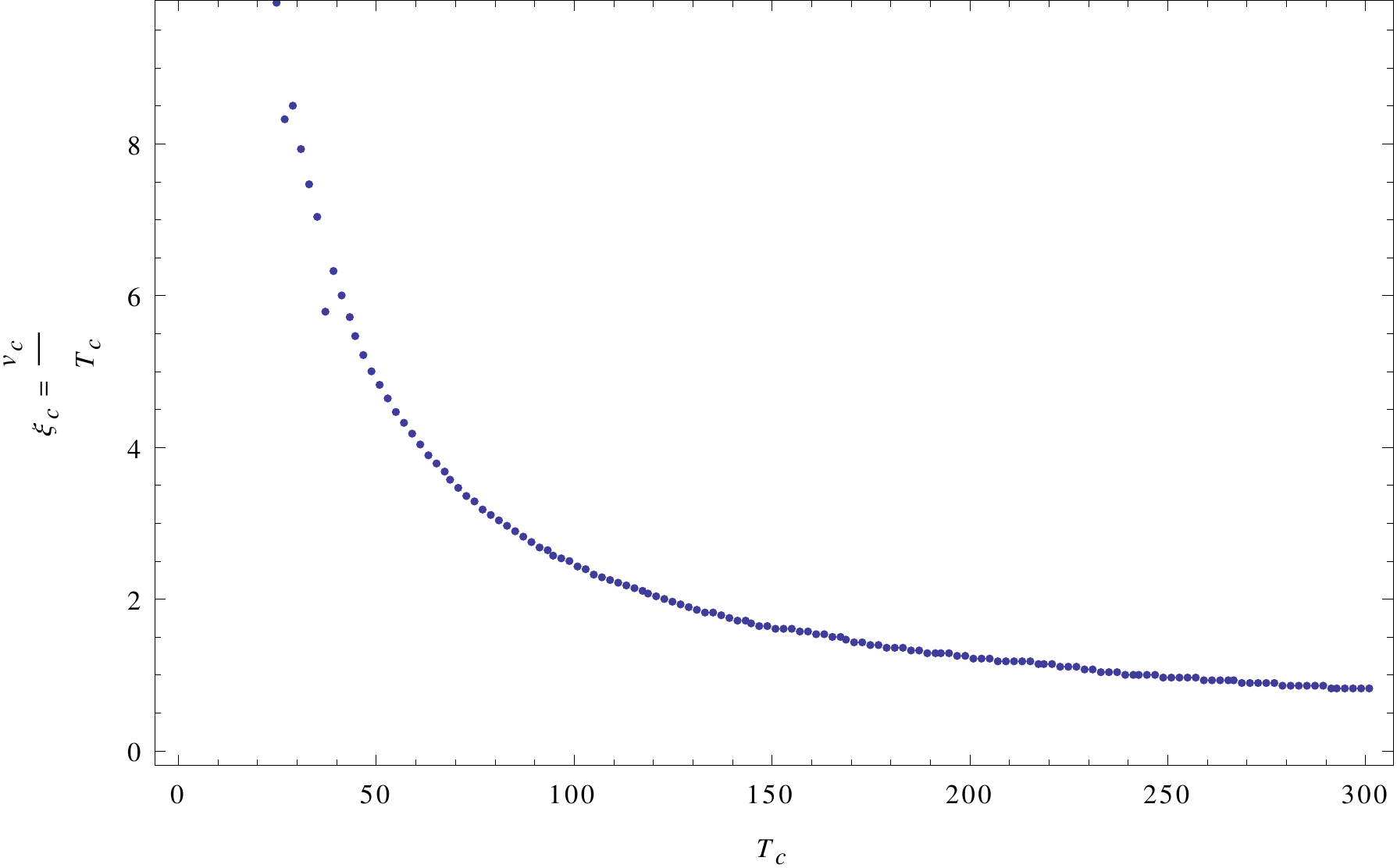}
\caption{\footnotesize The figure on the left is made with tan $\alpha$ = $\pi/4$ and on the right with tan $\alpha$=$\pi/8$. In Y-axis $\xi_{c}$ is plotted and it needs to be greater than or equal to 1 for SFOPT. Thus there's fair amount of points satisfying the condition.  
}\label{fig2}
\end{figure}
Here too a strong EWPT can occur. With a suitable critical temperature. 

Next we investigate if this can properly explain the di-photon excess. And the observed large cross-section as discussed in the previous section. Values of the input parameters are kept the same. The following graph is obtained. 

\begin{figure}[H]
\centering
 \includegraphics[width=75mm,height=75mm]{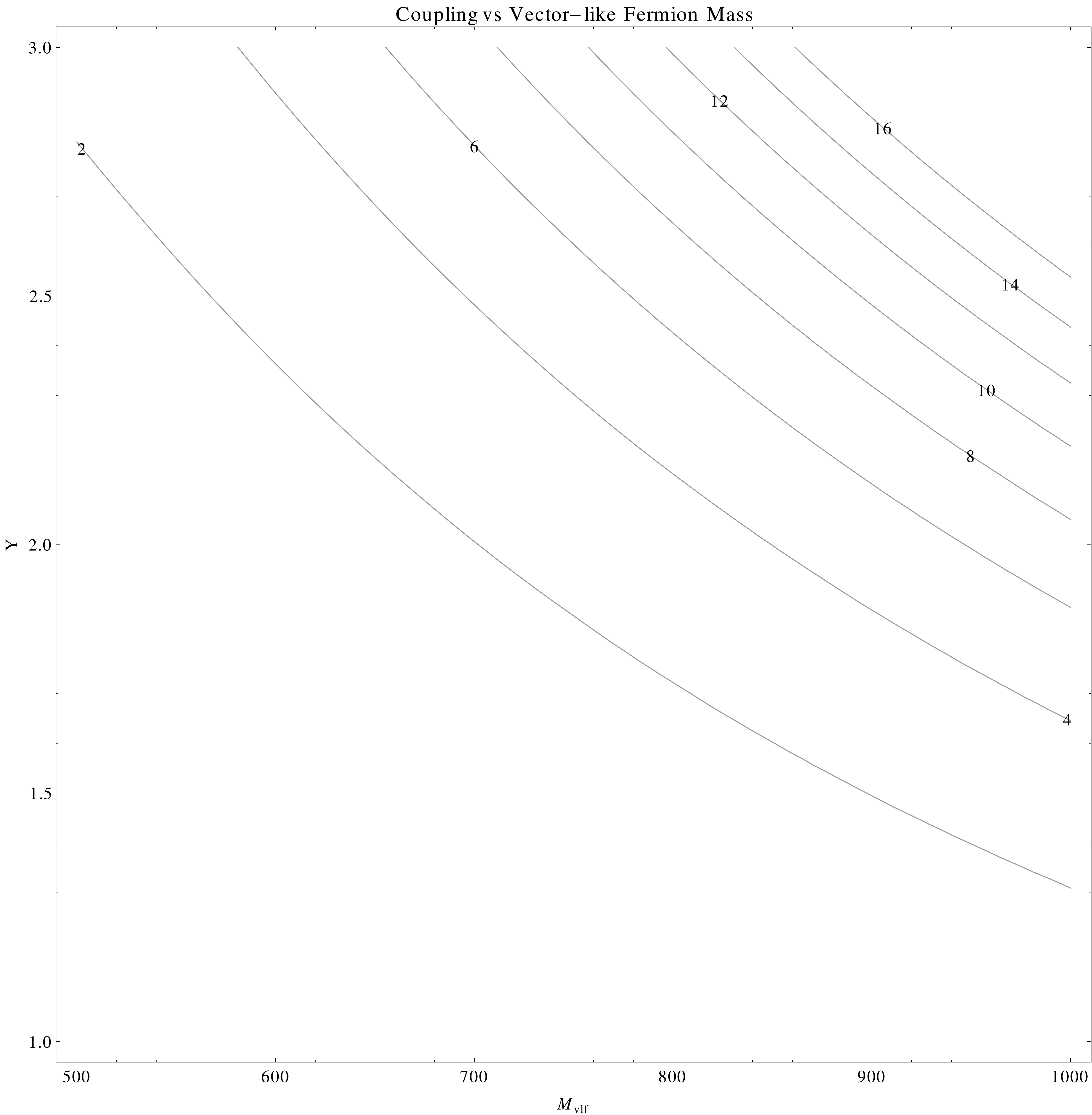}
 \includegraphics[width=75mm,height=75mm]{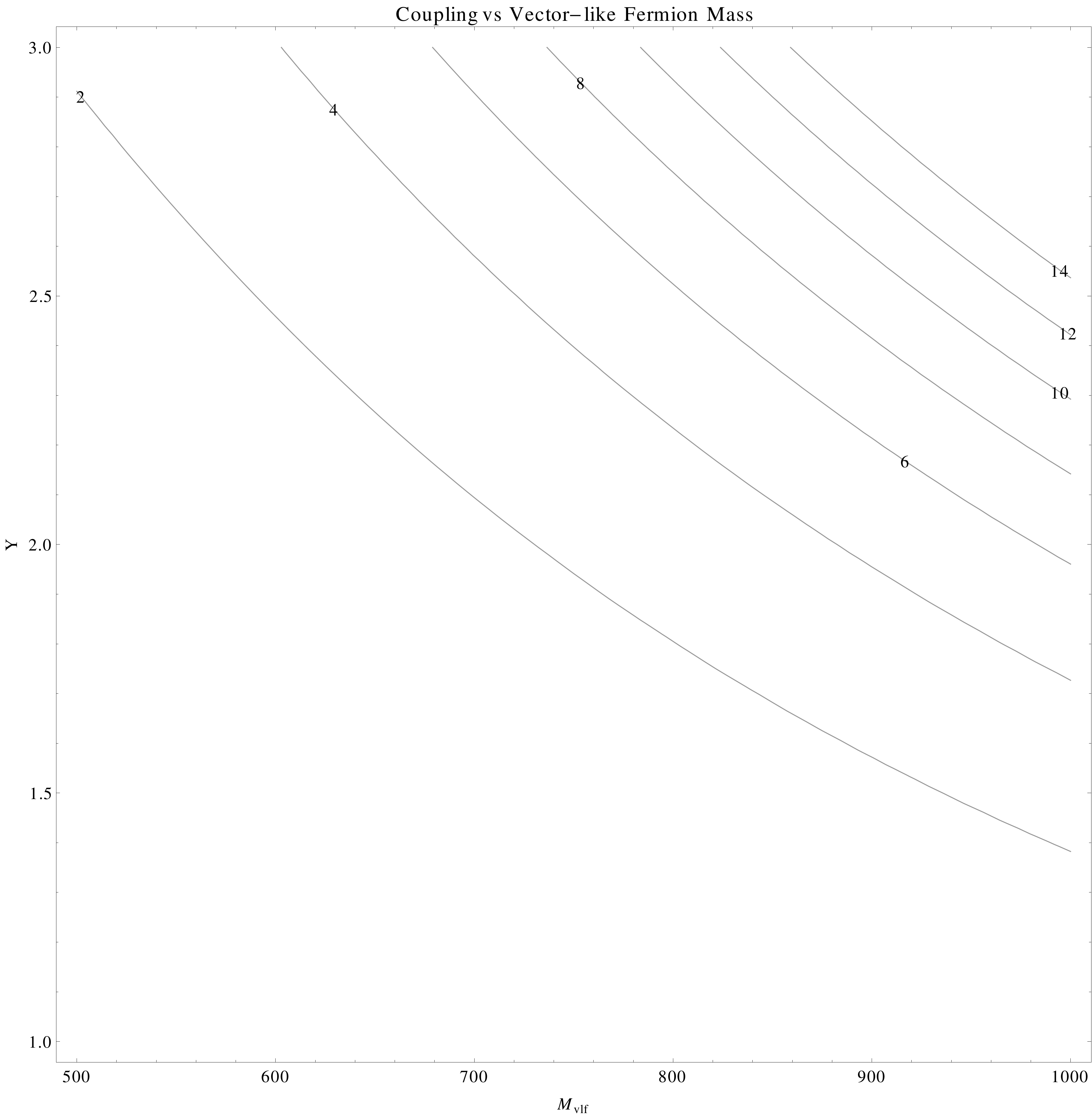}
 \includegraphics[width=75mm,height=75mm]{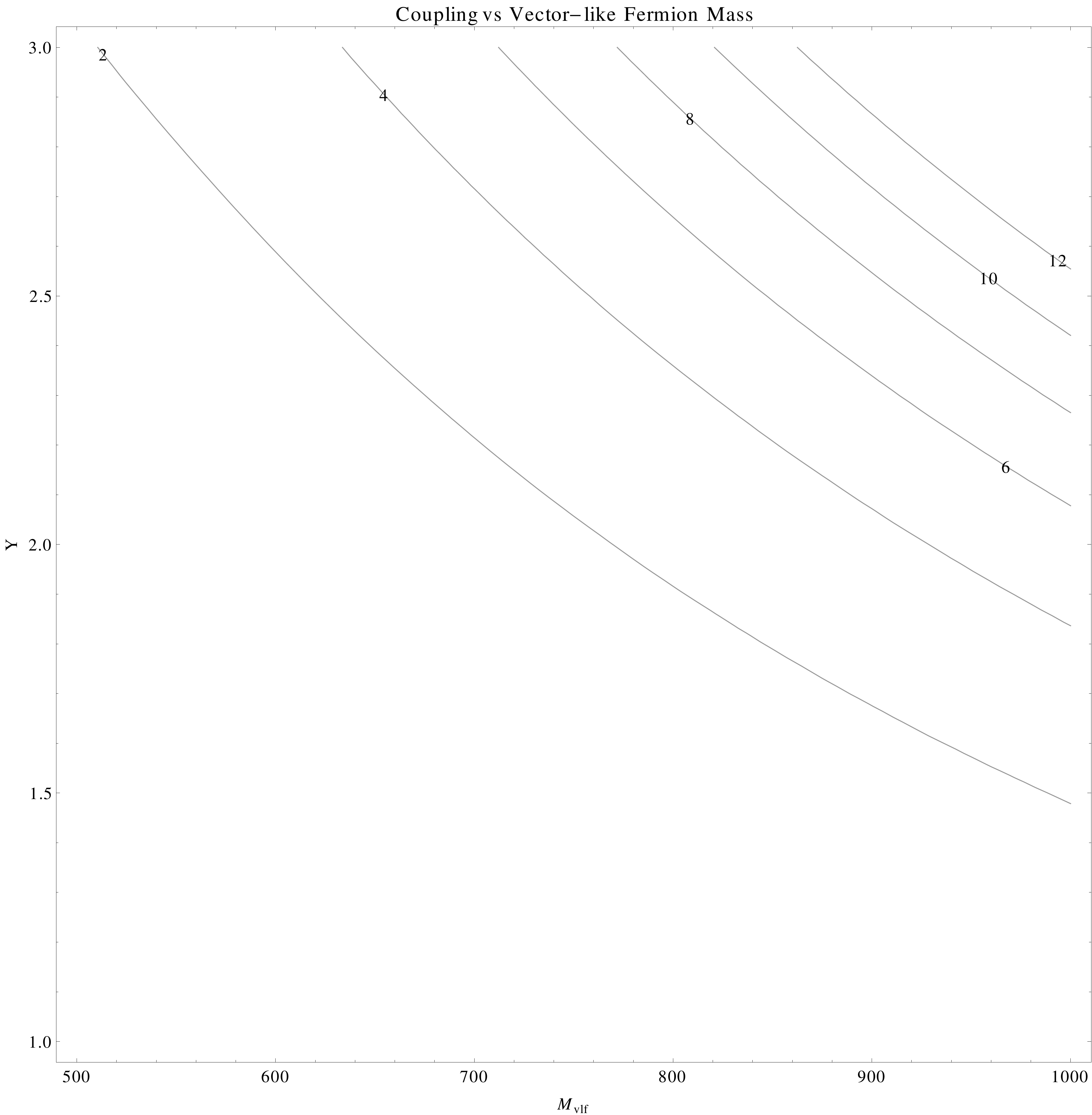}
 \includegraphics[width=75mm,height=75mm]{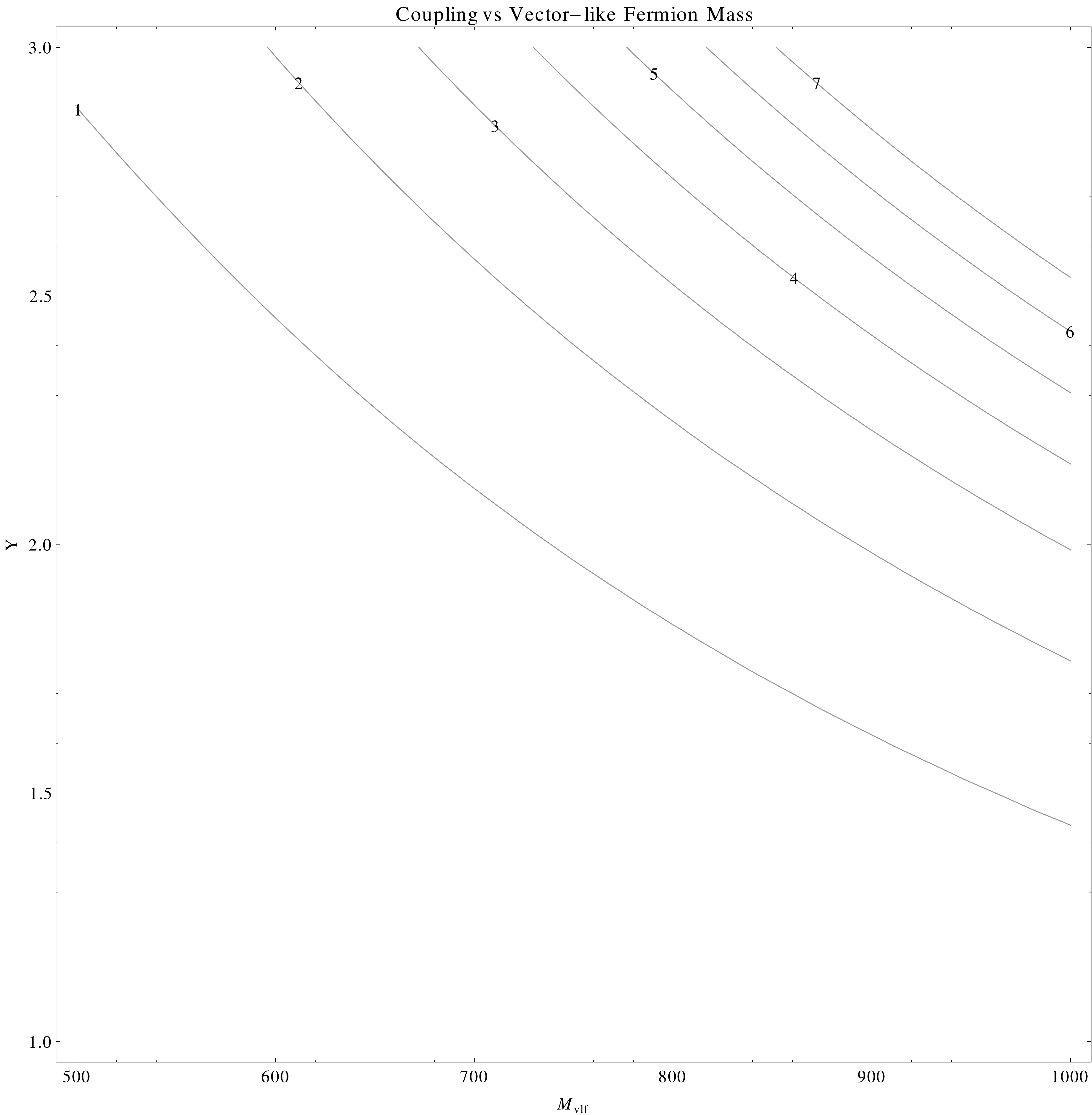}
       \caption{\footnotesize Plot of mass of the vector-like fermion vs its coupling to the 750 GeV scalar showing contours of cross-section in femto-barn units. Top-left fig is with $\alpha$=0.1, top-right is with $\alpha$=0.2, bottom-left is with $\alpha$=0.3 and the bottom-right fig is with $\alpha$=0.5 }\label{fig4}
\end{figure}
New particles (fermions or bosons) with a color and electric
charge and a large coupling to the 750 GeV scalar are needed to generate
an effective coupling of the scalar to gluons and photons. These new colored states
should not be too heavy, otherwise their couplings to the scalar must enter the non-
perturbative regime in order to explain the di-photon excess. As seen in the above figure, Y is of 2-2.5 for a vector-like fermion mass of 800-1000 GeV, is needed to get a cross-section around 7 $\mathrm{fb^{-1}}$ or so. One expects new colored colored states just around the corner provided this explanation to the di-photon excess is correct. 
We investigated the models and found that for a inert 750 GeV singlet scalar along with a vector-like fermion cannot be responsible for EWBG. While in the 2HDM it may happen. Again we need to add a vector-like fermion to accommodate the observed cross-section for di-photon excess. Choosing values within the previously studied constraints we could show that a strong first order EW phase transition is possible. 
\begin{figure}[H]
\centering
 \includegraphics[width=80mm]{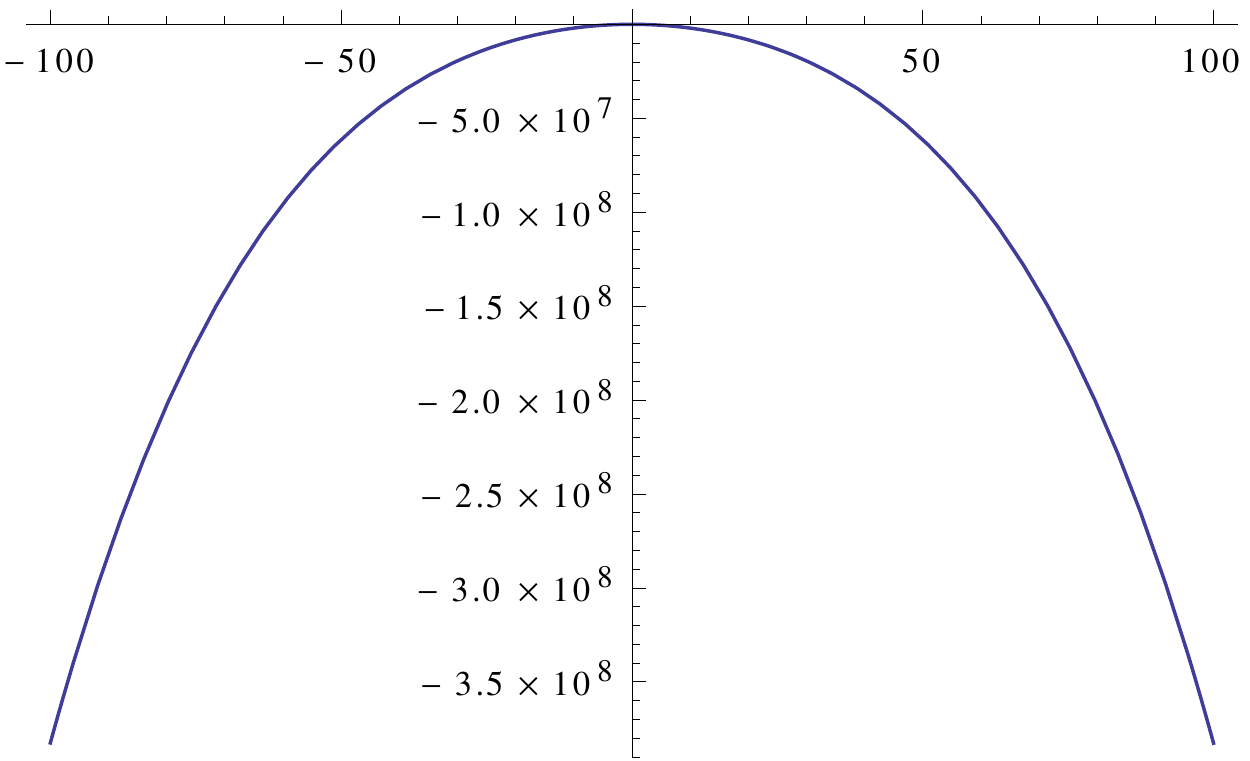}
 \includegraphics[width=80mm]{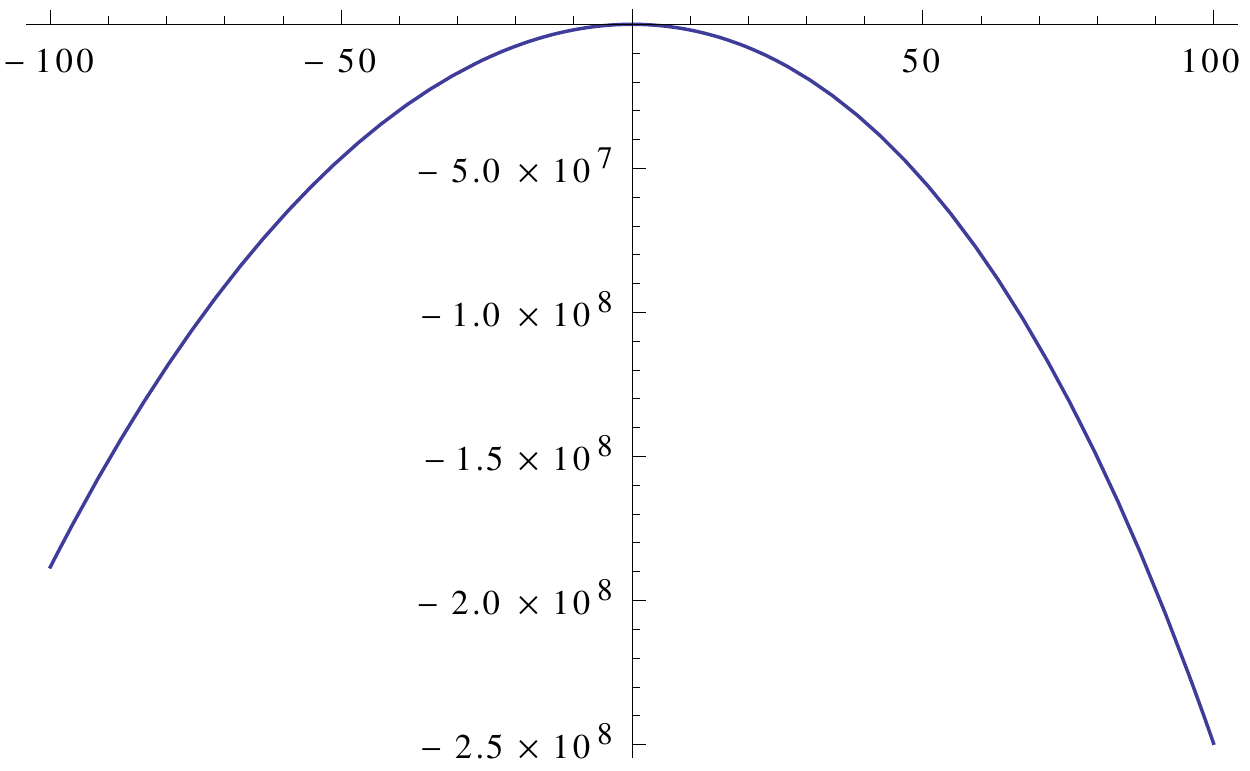}
 \includegraphics[width=80mm]{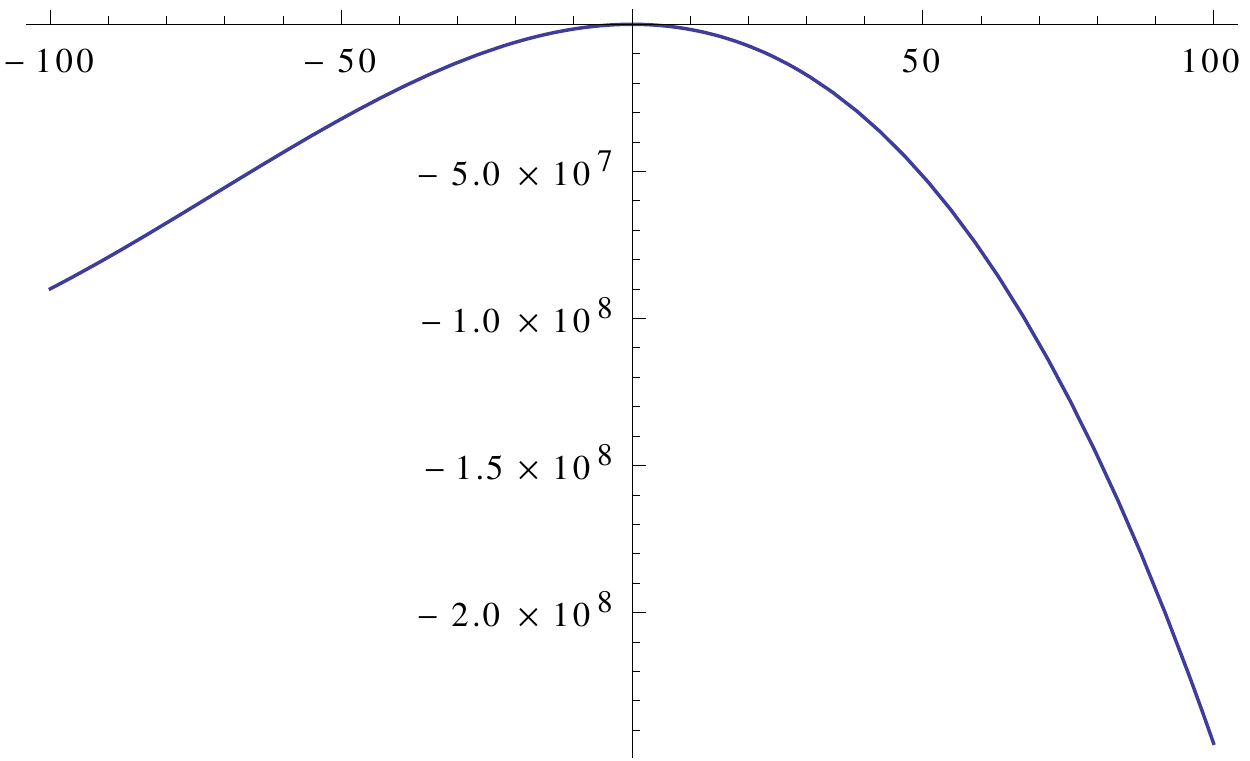}
 \includegraphics[width=80mm]{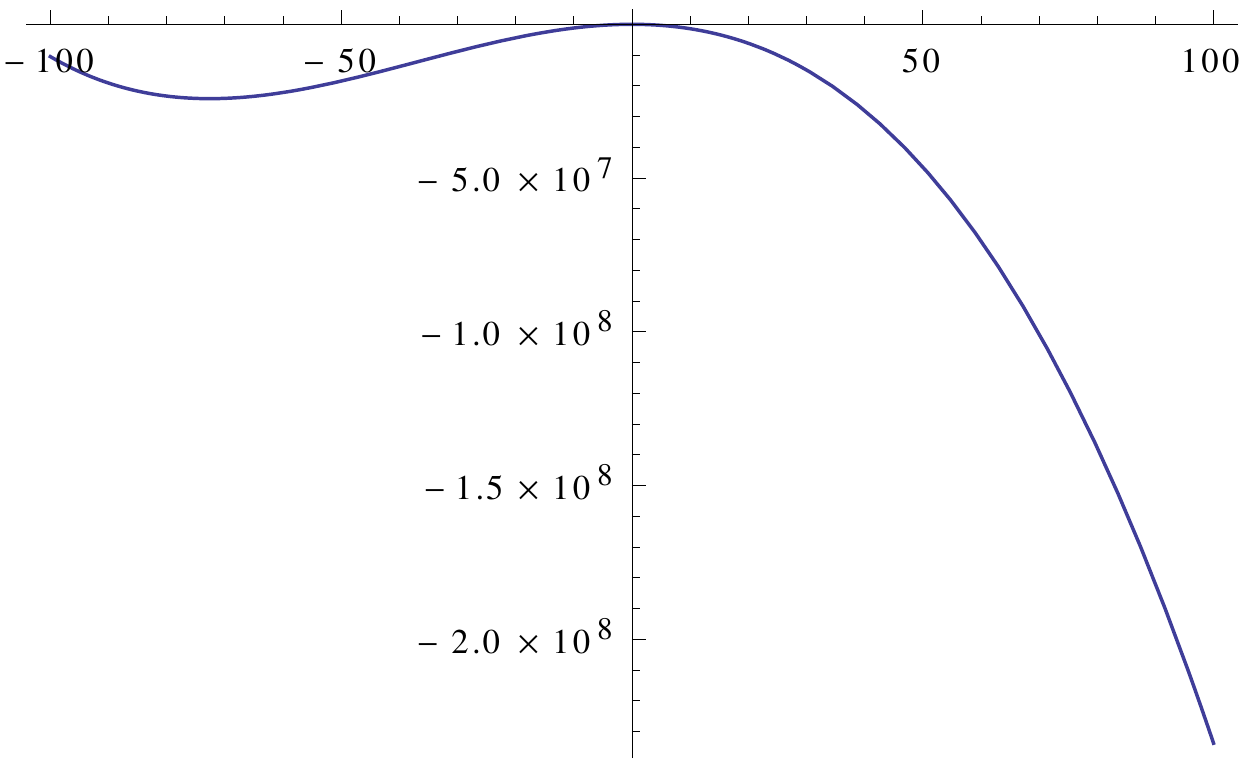}
 \includegraphics[width=80mm]{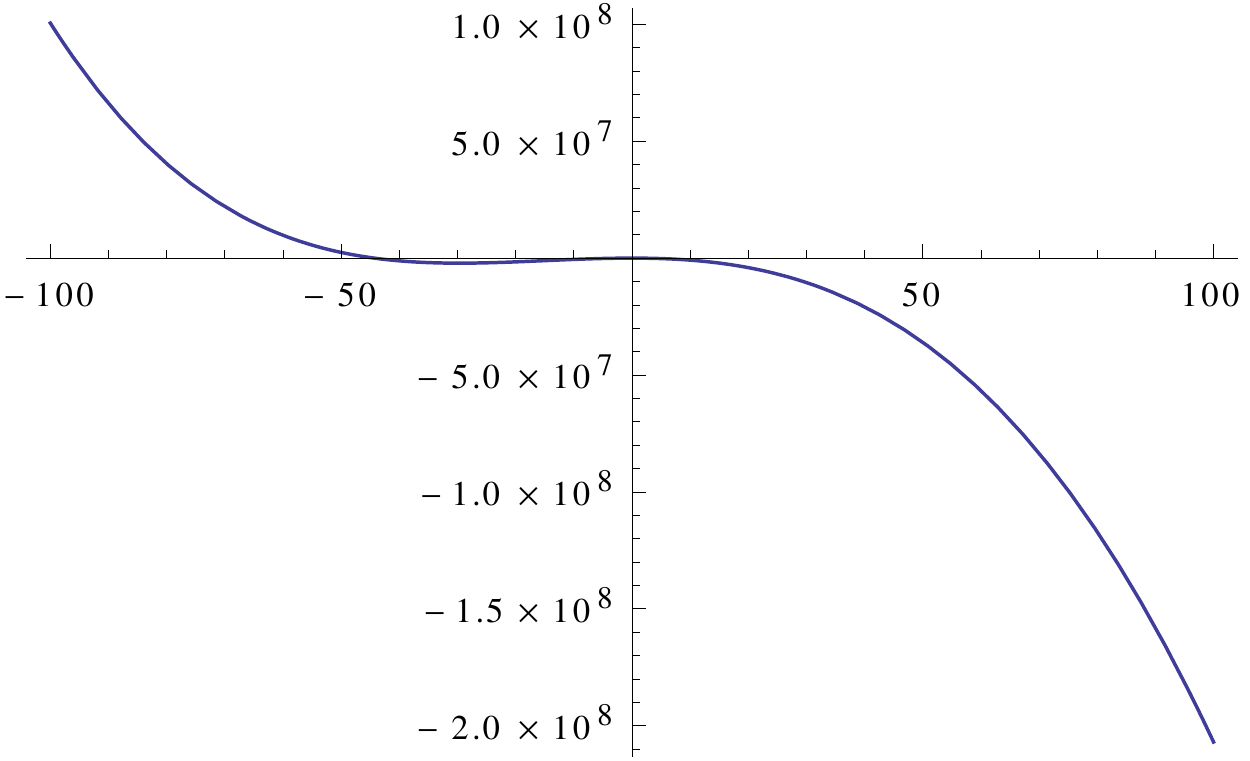}
 \includegraphics[width=80mm]{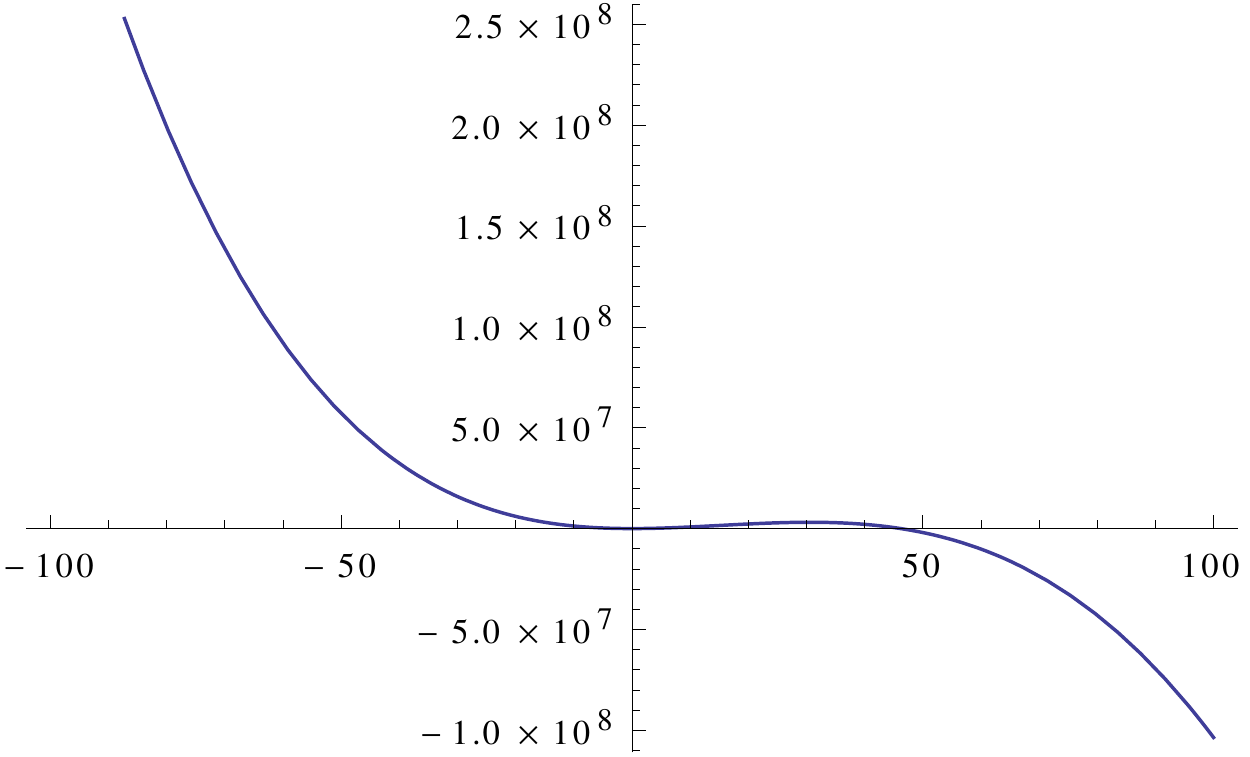}
 \includegraphics[width=80mm]{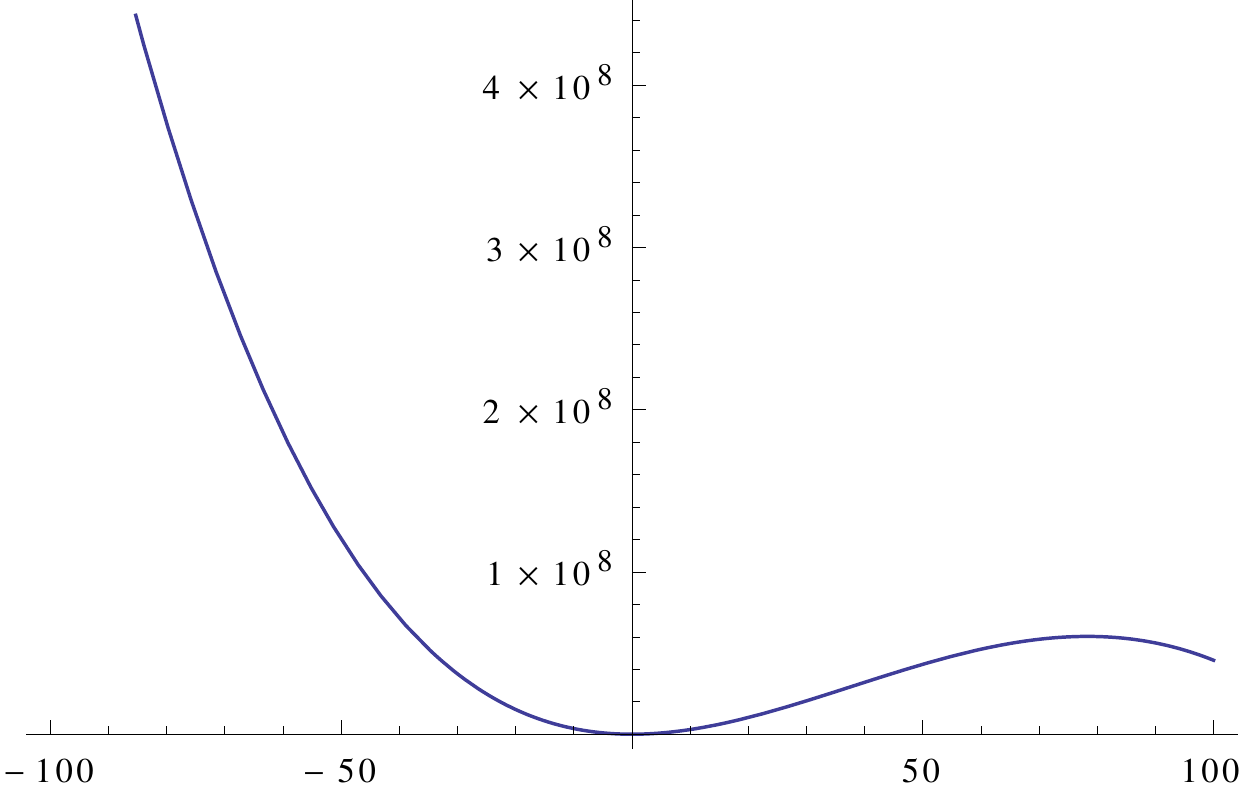}
 \includegraphics[width=80mm]{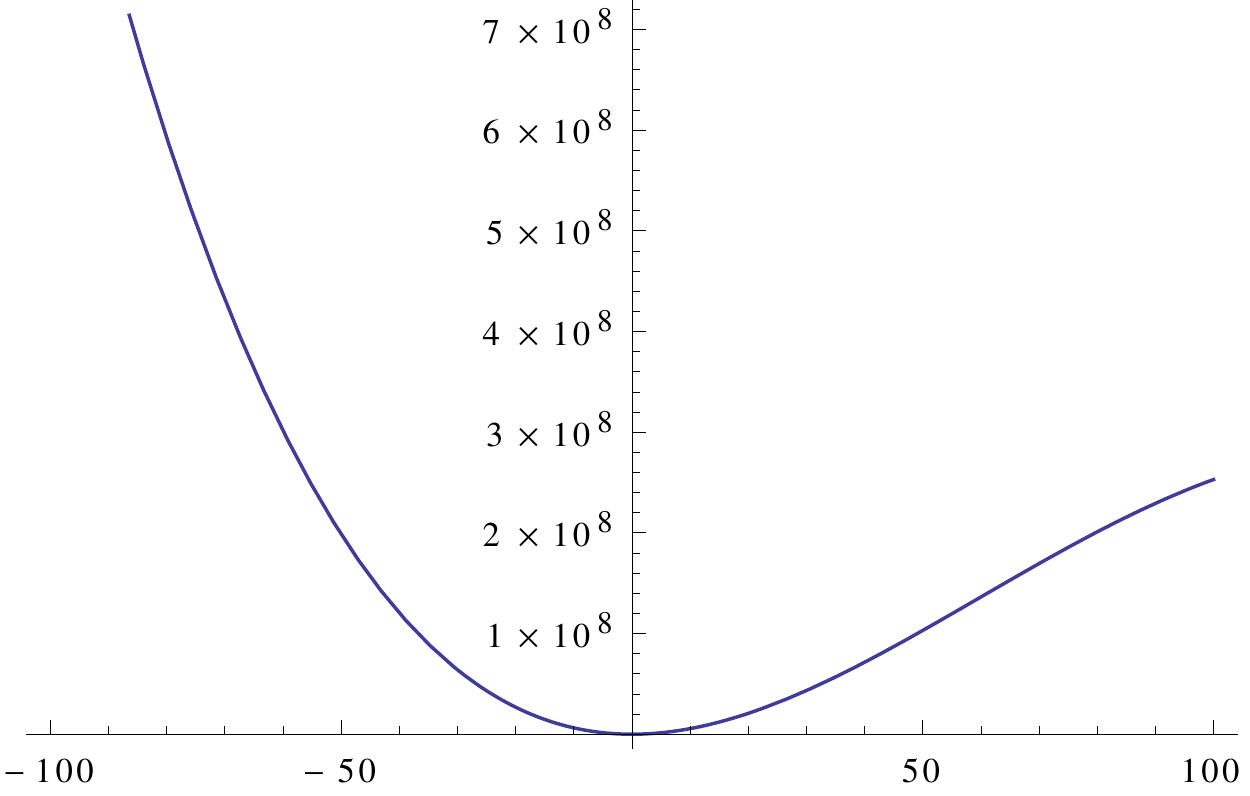}
       \caption{\footnotesize The figure shows the phase transition in the model. $V_{\textit{eff}}$ is plotted in Y-axis and VEV of the Higgs field in X-axis. As the temperature increases thermal effects leads to a transition from one minima to another. }\label{fig4}
\end{figure}
We see that the potential develops a thermal barrier. But serious issue is the applicability of the one-loop approximation and that of perturbation theory itself at finite temperatures. It is because the distribution function of bosons is large at low particle energies. At low temperatures interactions between the bosons is enhanced in the medium. Therefore this is known as the infrared problem. \cite{RubakovGourbunov}

\section{Other Models}
A glimpsing view on the existing literature on other scenarios for EW baryogenesis inspire us to investigate if the 750 GeV scalar can cause strong EWPT. Other than the 2HDM and the real scalar singlet there may a complex singlet extension model \cite{Barger:2008jx} or an abelian Higgs doublet model\cite{Jora:2015lsa}. Figure 8 in \cite{Ahriche:2015mea} shows an abundance of points satisfying the strong EWPT conditions and the scalar(s) being of 750 GeV mass. Besides the scale-invariant 2HDM which otherwise can cause strong 1st order phase transition as in \cite{Fuyuto:2015xsa}, it will be interesting to study with 750 GeV mass bound as done in the singlet model in this article. It however can explain the diphoton excess \cite{Antipin:2015kgh}. Slightly different scenario of 2-step baryogenesis \cite{Patel:2012pi,Inoue:2015pza} may also accommodate this new scalar resonance. Especially fig. 6 of the former looks promising as it is compatible with EDM bound too. While this article was being written, \cite{Lewicki:2016efe} appeared showing that the necessary condition
for the baryon asymmetry to avoid being 'washed out' after the phase transition may arise due the decoupling of fermion number violating processes, in to a modified expansion history caused due to this scalar dark matter. It will be interesting to study, with respect to this, the 750 GeV scalar resonance.
However a far more detailed study with respect to collider constraints need to be studied before we can make any further comments.

\section{Model-independent overview}
So we see that the veracity of the effective study of the Electroweak baryogenesis from the collider constraints vary on a model-to-model basis. However many individual models can shed light on the EWBG partially. As the LHC is running we would like to understand what classes of models may be consistent or at tension with with the data. 
Requirement of EWPT of strong first order means that two minima in the potential be separated by some barrier for a certain range of temperatures so that the system can undergo from one to the other by thermal fluctuations in addition to quantum fluctuations which has a decreasing probability with the increase in barrier. 
In addition to this barrier the phase transition requires EW sphaleron process to be out-of-equilibrium in the broken phase to ensure that the baryon asymmetry is not washed out. This condition is expressed as :\be  \frac{\nu_{c}}{T_{c}} \geq 1 \ee
Higgs data rules out EWBG as the mass of the Higgs being 125 GeV does not limit decay rate associated with the kinetic part of the Higgs action: the Nambu-Goldstone bosons. However if there's an loop-induced decay to di-photons then this problem can circumvented. 
The thermal barrier that we talked about earlier can be equivalently thought of as a thermal mass  $\approx h^{3}$ such that it can interplay with the $h^{2}$ and $h^{4}$ terms. Equation (17) is modified from model to model: 'e' becomes '$e + 6(\frac{\Lambda}{2})^{\frac{3}{2}}$' in the colored scalar model. And for the scalar singlet model by $e + n_{s} (\frac{\lambda_{m}}{4})^{\frac{3}{2}}$ where $n_{s}$ is the number of real scalar singlet degrees of freedom coupling to the Higgs.
For the 2HDM 'e' is replaced by '$e + 2 (\frac{\lambda_{3}}{2})^{\frac{3}{2}} + (\frac{\lambda_{3} + \lambda_{4} - \lambda_{5}}{2})^{\frac{3}{2}} + (\frac{\lambda_{3} + \lambda_{4} + \lambda_{5}}{2})^{\frac{3}{2}}$'.
So for the limit $\frac{\nu_{c}}{T_{c}}$ >> 1  we can reach the limit of large e. This means that the Higgs must have a large coupling with many light bosons. The reason that may impend this is the perturbativity bound and the Boltzman suppression for heavy bosons. Nonetheless one can keep on increasing the no. of bosonic degrees of freedom instead and they will contribute to the thermal loop and increase 'e' without treading the perturbativity bound. However in this case on also needs to take care of radiative corrections to $\lambda$ and must make the bosonic Higgs interactions loop-suppressed. e.g. \cite{Aad:2011dm}.
A second way is to take the limit of $\lambda$ to 0. This in a BSM means a light scalar. Which may be constrained by vacuum stability considerations \cite{EliasMiro:2011aa} and Higgs exotic decay. But it may be hidden in the large SM background.

A scenario may be such that the thermal barrier may arise from renormalizable tree-level Higgs interactions with the new scalar fields. This is given by the cubic term in the effective potential.
The strong PT condition  lead to $ \frac{\nu_{c}}{T_{c}} \approx \sqrt{\frac{4d}{\lambda\sqrt{1-\frac{\lambda m^{2}}{2e}}}}$. here the limits for increasing the fraction is d >> $\lambda$ which means increasing the coupling between the higgs and the light particles. This is similar to the e >> $\lambda$ of the previous case. Also if $\frac{\lambda m^{2}}{2e^{2}} \to 1$. The EW symmetric and broken vacua are degenerate and $T_{c}$ vanishes. This type of model is breifly discussed in \cite{Barger:2011vm}

\section{Comments}
In the work carried out it is understood that despite there being a fair possibility of the new 750 GeV scalar being responsible for strong first order phase transition and thus transferring the baryon asymmetry to our present universe CP violation is yet to be studied. Although some novel mechanisms exist which give an idea on CP-invariant asymmetry scenarios \cite{Hook:2015foa,Kobakhidze:2015zka},  most of the models introduce new sources of CP-violation. Not to mention the scalar resonance is yet to be confirmed at a  $ 5 \sigma$ level. So it may as well turn out to be but a statistical fluctuation. Finally the methods used in the present work may not be enough to ascertain certain claims impeccably. We have used elementary-level minimization techniques and algorithms. Parameters like temperatures were varied at 1 GeV step for the parameter space scan. It well may happen we have missed out on some minima lying in between. A more rigorous and comprehensive study will follow this. Similarly we plan to do other collider phenomenology studies like electroweak precision tests, etc. to constrain the properties of the scalar.   
Or missing energy search for the new fermions introduced. Other than covering a variety of models we also have in mind in putting this study on some existing theoretical framework like Supersymmetry.
At the end of the day with the present work it is being contemplated that the new scalar may have played a significant in early universe leading to observable effects at the scale that is being probed by the Large Hadron Collider.

\section{Acknowledgement}
I would like to thank Prof. Subhendra Mohanty, (Theory Division, Physical Research Laboratory, Ahmedabad) for his extraordinarily helpful suggestions and providing the opportunity to work in the institute during the winter. Discussions with Dr. Ujjal Kr. Dey, PRL and Dr. Saurabh Niyogi, IMSc. paved the way for a better understanding of the phenomena. Simultaneously I take the opportunity to convey my gratitude to Prof. James F. Libby and Prof. Suresh Govindarajan, I.I.T-Madras, Dr. Joydip Mitra, Scottish Church College, Kolkata and my friend Aritra Chowdhury for their support at different points of time in my career.

\end{document}